\documentclass[final,aps,amsmath,amssymb,nofootinbib,twocolumn]{revtex4-2}

\usepackage[utf8]{inputenc} 
\usepackage{graphicx}
\usepackage{hyperref} 
\usepackage{color} 
\usepackage{bm} 

\pdfoutput=1

\usepackage{mathtools}
\usepackage{slashed}
\usepackage{xcolor}
\usepackage{multirow}
\usepackage{gensymb}

\usepackage{hyperref}
\hypersetup{
    colorlinks=true,
    linkcolor=blue,
    citecolor=blue,
    urlcolor=blue
}

\usepackage[normalem]{ulem}

\usepackage{accents}

\usepackage{braket}

\allowdisplaybreaks

\usepackage{siunitx} 
\usepackage{xparse}

\usepackage{bbold}

\def\hat{\widehat}

\begin{document}

\title{Signals of nonrenormalizable Lorentz and CPT violation at the LHC}

\author{Enrico Lunghi}
\email{elunghi@iu.edu}
\affiliation{Physics Department, Indiana University, Bloomington, IN 47405, USA}
\affiliation{CERN, Theoretical Physics Department, Geneva, Switzerland}

\author{Nathaniel Sherrill}
\email{nathaniel.sherrill@itp.uni-hannover.de}
\affiliation{Institut f\"ur Theoretische Physik, Leibniz Universit\"at, Hannover, 30167, Germany}

\begin{abstract}
We examine nonrenormalizable Lorentz- and CPT-violating effective operators applied to the quark sector of the Standard Model. Using Drell-Yan events collected by the ATLAS and CMS Collaborations, 
several constraints are extracted from time-independent
modifications of the cross section on the $Z$-boson pole. The sensitivity to time-dependent modifications are also estimated by simulating a sidereal-time analysis. Our results suggest a dedicated search can
improve on constraints from deep inelastic scattering by up to three orders in magnitude. 
\end{abstract}

\maketitle 

\section{Introduction}
The search continues for violations of Lorentz and CPT invariance as signals of physics beyond the Standard Model (SM) and General Relativity (GR)~\cite{tables,Addazi:2021xuf}.
As fundamental spacetime symmetries, evidence for their violation would clearly suggest new physics. 
However, such violations need not be explicit in an underlying theory. Instead, they could appear as low-energy effective interactions coupling tensor-indexed vacuum expectation values (vevs)~\cite{Kostelecky:1994rn} or slowly varying fields to the SM or GR~\cite{Kostelecky:2002ca}. 
In the former case, one can view the discovery of the Higgs boson~\cite{ATLAS:2012yve,CMS:2012qbp} and the existence of a scalar vev as strong motivation to search for other background fields.
An example of the latter that has been drawing recent attention are ultralight dark matter candidates. As their collective behavior can be modeled as an oscillating background field, effective violations of Lorentz, CPT, and translation invariance can appear~\cite{Capozzi:2018bps,Gupta:2022qoq,Lambiase:2023hpq,Arguelles:2024cjj,Jiang:2024agx,Cordero:2024hjr}. 

Irrespective of any particular mechanism, and given the absence of a signal, employing a model-independent approach is a prudent strategy. The effective field theory (EFT) describing generic violations of fundamental spacetime symmetries is known as the Standard-Model Extension (SME)~\cite{Colladay:1996iz,Colladay:1998fq,Kostelecky:2003fs}. Treating the SM as an EFT and relaxing the condition of Lorentz invariance permits additional renormalizable and nonrenormalizable terms 
which form the SME action. Each term may be expressed as the contraction of an operator with a tensor-indexed coupling called an SME coefficient. Operators are constructed using the same quantum fields of the SM. By construction, the action is invariant under Lorentz coordinate transformations, referred to as observer transformations. As a consequence, the SME coefficients are tensors under these transformations.
Lorentz invariance is instead violated because transformations of the physical system, known as particle Lorentz transformations, act on the operators while leaving the unaffected the SME coefficients, resulting in a non-invariant contraction~\cite{Colladay:1996iz}. This pattern is precisely what would emerge if, for instance, Lorentz invariance were spontaneously broken.

A number of renormalizable SME coefficients involved in free propagation and electromagnetic interactions have been tightly constrained by a combination of phenomenological and experimental investigations~\cite{tables}. 
In contrast, effects associated to the lowest mass dimension ($d = 5,6$) nonrenormalizable terms, which have been completely enumerated~\cite{Bolokhov:2007yc,Kostelecky:2018yfa,Kostelecky:2020hbb}, remain largely unexplored. This is especially so for operators modifying the QCD sector of the SM.
As the sensitivity to nonrenormalizable operators increases with the probe scale, collider experiments provide an ideal setting for studying the associated effects on quarks and gluons. Some work in this direction has been carried out in Refs.~\cite{Kostelecky:2018yfa,Kostelecky:2019fse}, where $d=5$ operators involving light quarks were studied in the deep inelastic scattering (DIS) and Drell-Yan processes, assuming only photon exchange. These works were performed within the broader context of the Lorentz- and CPT-violating parton model~\cite{Kostelecky:2019fse}, which laid the groundwork for a recent search using five years of DIS data performed by the ZEUS Collaboration~\cite{ZEUS:2022msi}. Similar studies have been performed involving renormalizable SME effects on light and second-generation quarks~\cite{Kostelecky:2016pyx,Lunghi:2018uwj,Lunghi:2020hxn}, the $Z$ boson~\cite{Fu:2016fmf,Michel:2019tti}, and the top quark~\cite{Abazov:2012iu,Berger:2015yha,CMS:2024rcv,Belyaev:2024chj}.

In this paper, the effects of $Z$-boson exchange are consistently incorporated with the aforementioned $d=5$ operators. As discussed in Sec.~\ref{sec:ints}, the parity-violating nature of the weak interaction requires the presence of spin-independent and spin-dependent combinations of SME coefficients. 
The neutral-current Drell-Yan differential cross section is calculated and compared with LHC data.  We also use this opportunity to update previous constraints involving renormalizable ($d=4$) effects~\cite{Lunghi:2020hxn}. 

The remaining paper structure is as follows. In Sec.~\ref{sec:ints}, the interactions we consider are described and various definitions are introduced. In Sec.~\ref{sec:xsec}, the cross section is given and the time-independent and time-dependent contributions
are discussed. In Sec.~\ref{sec:numerical}, numerical results and constraints are presented and, where applicable, comparisons with previous results are made.

\section{Interactions}
\label{sec:ints}
The $d=5$ operators under consideration are invariant under $SU(2)\times U(1)$ electroweak gauge transformations and odd under CPT~\cite{Kostelecky:2020hbb}:
\begin{align}
\mathcal{L} =  
&
  -\frac{1}{2}a^{(5)\mu\alpha\beta}_{Q_i} \overline{Q}_i \gamma_\mu i D_{(\alpha}i  D_{\beta)} Q_i  \nonumber \\
&  -\frac{1}{2}a^{(5)\mu\alpha\beta}_{U_i} \overline{U}_i \gamma_\mu i D_{(\alpha}i  D_{\beta)} U_i  \\
&  -\frac{1}{2}a^{(5)\mu\alpha\beta}_{D_i} \overline{D}_i \gamma_\mu i D_{(\alpha}i  D_{\beta)} D_i  + \text{h.c.}, \nonumber
\label{eq:a5model}
\end{align}
where $a^{(5)\mu\alpha\beta}_{Q_i}$, 
$a^{(5)\mu\alpha\beta}_{U_i}$, and $a^{(5)\mu\alpha\beta}_{D_i}$
are SME coefficients, here assumed in units of GeV$^{-1}$.
The doublet and singlet fields have the conventional definitions 
\begin{align}
Q_i = \begin{pmatrix} u_i \\ d_i \end{pmatrix}_L, \quad U_i = \left(u_i\right)_R, \quad D_i = \left(d_i\right)_R, 
\end{align}
where $i = 1, 2, 3$ denotes flavors $u_i = (u, c, t)$ and $d_i = (d, s, b)$,
and $i D_{(\alpha}i  D_{\beta)}$ 
denotes the symmetrized product 
of conventional gauge-covariant derivatives. 
Using the freedom to redefine the fields shows that only the totally symmetric and traceless components of the SME coefficients are observable. These are denoted as $a_{\text{S}Q_i}^{(5)\mu\alpha\beta}, a_{\text{S}U_i}^{(5)\mu\alpha\beta},$ and $a_{\text{S}D_i}^{(5)\mu\alpha\beta}$, where each type has 16 independent components~\cite{Ding:2016lwt,Edwards:2019lfb,Foster:2016uui}. 
Assuming these redefinitions have been performed, the coefficients expressed in the mass eigenstate basis are 
\begin{align}
\label{eq:quarkcoeffs1}
a_{\text{S}u_i}^{(5)\mu\alpha\beta} &= (a_{\text{S}Q_i}^{(5)\mu\alpha\beta}+a_{\text{S}U_i}^{(5)\mu\alpha\beta})/2\;, \\
a_{\text{S}d_i}^{(5)\mu\alpha\beta} &= (a_{\text{S}Q_i}^{(5)\mu\alpha\beta}+a_{\text{S}D_i}^{(5)\mu\alpha\beta})/2 \;, \\
b_{\text{S}u_i}^{(5)\mu\alpha\beta} &= (a_{\text{S}Q_i}^{(5)\mu\alpha\beta}-a_{\text{S}U_i}^{(5)\mu\alpha\beta})/2 \; , \\
b_{\text{S}d_i}^{(5)\mu\alpha\beta} &= (a_{\text{S}Q_i}^{(5)\mu\alpha\beta}-a_{\text{S}D_i}^{(5)\mu\alpha\beta})/2 \;,\label{eq:quarkcoeffs4}
\end{align}
where $a_{\text{S}u_i}^{(5)\mu\alpha\beta} -a_{\text{S}d_i}^{(5)\mu\alpha\beta} = b_{\text{S}d_i}^{(5)\mu\alpha\beta} -b_{\text{S}u_i}^{(5)\mu\alpha\beta}$. We use this constraint for the first two generations to eliminate $b_{\text{S}d_i}^{(5)\mu\alpha\beta}$. For the third generation, due to the absence of $a_{\text{S}U_3}^{(5)\mu\alpha\beta}$, we consider $a_{\text{S}b}^{(5)\mu\alpha\beta}$ and $b_{\text{S}b}^{(5)\mu\alpha\beta}$. In the mass eigenstate basis, the $a^{(5)}$-type and $b^{(5)}$-type coefficients are associated with spin-independent and spin-dependent effects, respectively. 
\section{Cross section}
\label{sec:xsec}
Following the approach of leading-order factorization in the presence of Lorentz violation~\cite{Kostelecky:2019fse,Lunghi:2020hxn}, we find that the differential cross section is given by 
\begin{align}
\label{DYsigmadQ2} \frac{d\sigma}{dQ^2}    =\; &    
\frac{4\pi\alpha^2}{3 N_c}
\sum_f \Bigg[\frac{e_f^2}{2Q^4} \nonumber \\ 
& \hskip-0.1cm + \frac{1-m_Z^2/Q^2}{(Q^2 -m_Z^2)^2 + m_Z^2\Gamma_Z^2} \frac{1-4\sin^2\theta_W}{4\sin^2\theta_W\cos^2\theta_W}e_fg_{fL} \nonumber \\
& \hskip-0.1cm  + \frac{1}{(Q^2 -m_Z^2)^2 
+ m_Z^2\Gamma_Z^2}\frac{1+(1-4\sin^2\theta_W)^2}{32\sin^4\theta_W\cos^4\theta_W}g_{fL}^2\Bigg] \nonumber \\
&\hskip-0.1cm \times 
\int_{\tau}^1 dx\frac{\tau}{x}\hat{\sigma}_{f} 
\left(x,\tau,a_{\text{S}f_L}^{(5)}\right) + (L\rightarrow R),
\end{align}
where $f=\{u_i,d_i\} = \{u,d,s,c,b\}$ and the three terms in brackets correspond to photon, photon-$Z$ interference, and $Z$ exchange. The integrand is 
\begin{align}
\label{sigmaprime}
\hat{\sigma}_{f}\left(x,\tau, a_{\text{S}f_L}^{(5)} \right) = \; & 
\left[ 1 + A_{\text{S}}\left(x,\tau, a_{\text{S}f_L}^{(5)}\right) \right] f_{\text{S}f}(x,\tau/x) \nonumber \\
& +A'_{\text{A}}\left(x,\tau, a_{\text{S}f_L}^{(5)}\right)f_{\text{A}f}(x,\tau/x)  \nonumber \\
& \vphantom{\frac{\int}{\int}}+ A_{\text{A}}\left(x,\tau, a_{\text{S}f_L}^{(5)}\right)f'_{\text{A}f}(x,\tau/x)  \; , 
\end{align}
where the symmetric and antisymmetric quark 
$f_f(x)$ and antiquark $f_{\bar f}(x)$ 
products of parton distribution functions (PDFs) are defined as
\begin{align}
f_{\text{S}f}(x,\tau/x) \equiv \; & f_f(x)f_{\bar f}(\tau/x) + f_f(\tau/x)f_{\bar f}(x), \\
f_{\text{A}f}(x,\tau/x) \equiv  \; & f_f(x)f_{\bar f}(\tau/x) - f_f(\tau/x)f_{\bar f}(x),  \label{eq:fA} \\
f'_{\text{A}f}(x,\tau/x)  \equiv  \; &f_f(x)f'_{\bar f}(\tau/x) - f'_f(\tau/x)f_{\bar f}(x),
\label{pdfdefs}
\end{align}
with $f_f'(\tau/x)$ the derivative of the PDF evaluated at $\tau/x$. The prefactor
functions are expressed in terms of the chiral coefficients
$a_{\text{S}u_{iL}}^{(5)\mu\alpha\beta} = 
a_{\text{S}d_{iL}}^{(5)\mu\alpha\beta} \equiv  
a_{\text{S}Q_i}^{(5)\mu\alpha\beta}$,
$a_{\text{S}u_{iR}}^{(5)\mu\alpha\beta} \equiv 
a_{\text{S}U_i}^{(5)\mu\alpha\beta}$, 
$a_{\text{S}d_{iR}}^{(5)\mu\alpha\beta} 
\equiv a_{\text{S}D_i}^{(5)\mu\alpha\beta}$ 
and given by 
\begin{widetext}
\begin{align}
A_{\text{S}}\left(x,\tau, a_{\text{S}f_L}^{(5)}\right)  
= \; & \frac{2}{s}(x+\tau/x)a_{\text{S}f_L}^{(5)\mu\alpha\beta} \left(p_{1\mu}p_{1\alpha}p_{2\beta} + p_{1\mu}p_{2\alpha}p_{2\beta}\right) \nonumber \\
=  \; & (x + \tau/x) E_p\left(a_{\text{S}f_L}^{(5)000} - a_{\text{S}f_L}^{(5)033} \right)  ,\label{AS} \\
A_{\text{A}}\left(x,\tau, a_{\text{S}f_L}^{(5)}\right) 
=\; & -\frac{1}{2sx}a_{\text{S}f_L}^{(5)\mu\alpha\beta}\left[ (x-\tau/x)(x+\tau/x)^2\left(p_{1\mu}p_{1\alpha}p_{1\beta} + p_{1\mu}p_{1\alpha}p_{2\beta} + (p_1\leftrightarrow p_2)\right) \right. \nonumber\\
&\left. 
+ (x-\tau/x)\left(x^2 + (\tau/x)^2\right)
\left(p_{1\mu}p_{1\alpha}p_{1\beta} - p_{1\mu}p_{1\alpha}p_{2\beta} 
+ (p_1\leftrightarrow p_2)\right) \right] \nonumber \\
= \; & -\frac{1}{x}E_p \left[\tfrac{1}{2}(x-\tau/x)(x+\tau/x)^2\left(a_{\text{S}f_L}^{(5)000}  + a_{\text{S}f_L}^{(5)033}\right) + a_{\text{S}f_L}^{(5)033}(x-\tau/x)\left(x^2+(\tau/x)^2\right)  \right], \label{AA}\\
A'_{\text{A}}\left(x,\tau, a_{\text{S}f_L}^{(5)}\right) 
= \; & \frac{1}{2sx^3} a_{\text{S}f_L}^{(5)\mu\alpha\beta} \left[\left(x^4 -2\tau x^2  + 3\tau^2\right)\left(p_{1\mu}p_{1\alpha}p_{1\beta} - p_{1\mu}p_{1\alpha}p_{2\beta} + (p_1\leftrightarrow p_2)\right) \right. \nonumber \\
&\left. - \left(x^2 - 3\tau\right) \left(x^2  + \tau\right)\left(p_{1\mu}p_{1\alpha}p_{1\beta} 
+ p_{1\mu}p_{1\alpha}p_{2\beta} + (p_1\leftrightarrow p_2)\right)\right]\nonumber \\
 = \; &  \; \frac{1}{2x^3}E_p \left[2\left(x^4 -2\tau x^2 + 3\tau^2\right)a_{\text{S}f_L}^{(5)033} - \left(x^2 - 3\tau\right)\left(x^2  + \tau\right)(a_{\text{S}f_L}^{(5)000} + a_{\text{S}f_L}^{(5)033})\right] . \label{APA}
\end{align}
\end{widetext}
The first lines in each of Eqs.~\eqref{AS}--\eqref{APA} display the functions in observer-invariant form. Each of the second lines is the value of the function evaluated in the laboratory frame, where only two coefficients $a_{\text{S}fL,R}^{000}$ and $a_{\text{S}fL,R}^{033}$ appear. This mirrors the appearance of only the renormalizable coefficients $c_{fL,R}^{00}$ and $c_{fL,R}^{33}$ in the same observable from Ref.~\cite{Lunghi:2020hxn}. 
In the limit of photon exchange, the results of Ref.~\cite{Kostelecky:2019fse} are reproduced, and setting the SME coefficients to zero yields the leading-order SM result.

There are two important points regarding the differential cross section in Eq.~\eqref{DYsigmadQ2} that are worth emphasizing. First, the SME coefficients have negative mass dimension and always appear in combination with the proton energy $E_p$, implying larger effects at higher collision energies.
Our calculations assume the SME coefficients are small perturbations; therefore,  
in order to guarantee the smallness of second-order contributions, we require $a_{\text{S}f_{L,R}}^{(5)}  E_p \lesssim 1$. 
Second, the contributions involving the antisymmetric combinations of PDFs, Eqs.~(\ref{eq:fA}) and (\ref{pdfdefs}), vanish for the second and third generation due to the near equality of quark and antiquark sea PDFs. Since the largest contributions stem from terms proportional to $f_{\text{A}f}^\prime$~\cite{Kostelecky:2019fse}, we expect lower sensitivity to the SME coefficients for second- and third-generation quarks.

It is common convention to re-express the laboratory-frame SME coefficients in terms of SME coefficients with indices $T, X, Y, Z$ defined in the Sun-centered frame (SCF)~\cite{Kostelecky:2002hh,Bluhm:2001rw,Bluhm:2003un}. After doing so, the differential cross section contains time-independent and time-dependent Lorentz-violating effects. The latter are given by periodic trigonometric functions controlled by first and second harmonics of the Earth's sidereal frequency $\omega_\oplus \approx 2\pi/(\text{23h:56min:4s})$. As the coefficients have three Lorentz indices, third-order harmonics could in principle appear. They do not for this particular observable due to the symmetric integration over parton momenta~\cite{Kostelecky:2019fse}.

The explicit expressions for the laboratory coefficients in terms of the SCF coefficients are $a_{\text{S}f_{L,R}}^{(5)000} = a_{\text{S}f_{L,R}}^{(5)TTT}$ and 
\begin{widetext}
\begin{align}
\label{a033}
a_{\text{S}f_{L,R}}^{(5)033} =&
\frac{1}{2}(a_{\text{S}f_{L,R}}^{(5)TXX}+a_{\text{S}f_{L,R}}^{(5)TYY})\left(\cos^2\chi\sin^2\psi + \cos^2\psi\right) 
+ a_{\text{S}f_{L,R}}^{(5)TZZ}\sin^2\chi\sin^2\psi \\
& - 2a_{\text{S}f_{L,R}}^{(5)TXZ}\sin\chi\sin\psi
\left[\cos\chi\sin\psi\cos(\omega_{\oplus}T_{\oplus} )
+ \cos\psi\sin(\omega_{\oplus}T_{\oplus})\right]
\nonumber\\
& - 2a_{\text{S}f_{L,R}}^{(5)TYZ}\sin\chi\sin\psi
\left[\cos\chi\sin\psi\sin(\omega_{\oplus}T_{\oplus})
- \cos\psi\cos(\omega_{\oplus}T_{\oplus})\right] \nonumber\\
& + a_{\text{S}f_{L,R}}^{(5)TXY}\left[(\cos^2\chi\sin^2\psi - \cos^2\psi)\sin(2\omega_\oplus T_\oplus) - \cos\chi\sin(2\psi)\cos(2\omega_\oplus T_\oplus)\right] \nonumber \\
& +\frac{1}{2}(a_{\text{S}f_{L,R}}^{(5)TXX}-a_{\text{S}f_{L,R}}^{(5)TYY})\left[(\cos^2\chi\sin^2\psi - \cos^2\psi)\cos(2\omega_\oplus T_\oplus) + \cos\chi\sin(2\psi)\sin(2\omega_\oplus T_\oplus)\right], \nonumber
\end{align}
\end{widetext}
where $\chi$ and $\psi$ are the laboratory colatitude and beamline orientation north of east, respectively. The convention used for the directions of the laboratory cartesian axes is consistent with Ref.~\cite{ATLAS:2019zci}. Expressed in this form, it is clear that the SCF coefficients with indices $TTT$, $TXX + TYY$, and $TZZ$ are associated with time-independent effects, while those with indices $TXZ$ and $TYZ$ ($TXY$ and $TXX-TYY$) introduce time-dependent effects involving the local sidereal time $T_\oplus$ and the first (second) sidereal harmonics.

\def\arraystretch{1.2}
\begin{table}[t]
\centering
\begin{tabular}{|c||c|c|}\hline\hline
coefficient [$\text{GeV}^{-1}$]   & lower & upper \\ \hline
$a_{\text{S}u}^{(5)TTT}$ & $-4.1\times 10^{-8}$ & $1.2\times 10^{-7}$   \\
$a_{\text{S}u}^{(5)TZZ}$ & $-2.2\times 10^{-7}$ & $7.1\times 10^{-8}$   \\\hline
$b_{\text{S}u}^{(5)TTT}$ & $-5.4\times 10^{-8}$ & $1.6\times 10^{-7}$   \\
$b_{\text{S}u}^{(5)TZZ}$ & $-2.8\times 10^{-7}$ & $9.3\times 10^{-8}$   \\\hline
$a_{\text{S}d}^{(5)TTT}$ & $-2.2\times 10^{-6}$ & $6.5\times 10^{-6}$   \\
$a_{\text{S}d}^{(5)TZZ}$ & $-1.1\times 10^{-5}$ & $3.7\times 10^{-6}$   \\\hline\hline
$a_{\text{S}c}^{(5)TTT}$ & \textcolor{gray}{$-1.5\times 10^{-3}$} & \textcolor{gray}{$4.5\times 10^{-3}$}   \\
$a_{\text{S}c}^{(5)TZZ}$ & \textcolor{gray}{$-1.6\times 10^{-3}$} & \textcolor{gray}{$4.8\times 10^{-3}$}   \\\hline
$b_{\text{S}c}^{(5)TTT}$ & \textcolor{gray}{$-1.6\times 10^{-3}$} & \textcolor{gray}{$4.9\times 10^{-3}$}   \\
$b_{\text{S}c}^{(5)TZZ}$ & \textcolor{gray}{$-1.7\times 10^{-3}$} & \textcolor{gray}{$5.3\times 10^{-3}$}   \\\hline
$a_{\text{S}s}^{(5)TTT}$ & \textcolor{gray}{$-3.0\times 10^{-2}$} & \textcolor{gray}{$9.1\times 10^{-2}$}   \\
$a_{\text{S}s}^{(5)TZZ}$ & \textcolor{gray}{$-3.2\times 10^{-2}$} & \textcolor{gray}{$9.7\times 10^{-2}$}   \\\hline\hline
$a_{\text{S}b}^{(5)TTT}$ & \textcolor{gray}{$-9.8\times 10^{-3}$} & \textcolor{gray}{$3.0\times 10^{-2}$}   \\
$a_{\text{S}b}^{(5)TZZ}$ & \textcolor{gray}{$-1.0\times 10^{-2}$} & \textcolor{gray}{$3.2\times 10^{-2}$}   \\\hline
$b_{\text{S}b}^{(5)TTT}$ & \textcolor{gray}{$-1.0\times 10^{-2}$} & \textcolor{gray}{$3.2\times 10^{-2}$}   \\
$b_{\text{S}b}^{(5)TZZ}$ & \textcolor{gray}{$-1.1\times 10^{-2}$} & \textcolor{gray}{$3.4\times 10^{-2}$}   \\\hline\hline
\end{tabular}
\caption{Constraints on the $d=5$ coefficients that do not induce sidereal-time dependence, in units of $\text{GeV}^{-1}$. Constraints in gray do not satisfy the condition of perturbativity (coefficient $\lesssim E_p^{-1} \simeq 10^{-4}$\;GeV$^{-1}$). \label{tab:MZindependent}}
\end{table}

\def\arraystretch{1.2}
\begin{table}[t]
\centering
\begin{tabular}{|c||c|c|}\hline\hline
coefficient \vphantom{$\Big($} & lower & upper \\ \hline
$c_{u}^{TT}$ & $-4.3\times 10^{-2}$ & $1.4\times 10^{-2}$   \\
$c_{u}^{ZZ}$ & $-4.3\times 10^{-4}$ & $1.3\times 10^{-3}$   \\\hline
$d_{u}^{TT}$ & $-5.6\times 10^{-2}$ & $1.8\times 10^{-2}$   \\
$d_{u}^{ZZ}$ & $-5.6\times 10^{-4}$ & $1.7\times 10^{-3}$   \\\hline
$c_{d}^{TT}$ & \textcolor{gray}{$-2.0$} & $6.6\times 10^{-1}$   \\
$c_{d}^{ZZ}$ & $-1.9\times 10^{-2}$ & $5.9\times 10^{-2}$   \\\hline\hline
$c_{c}^{TT}$ & $-4.5\times 10^{-1}$ & $1.5\times 10^{-1}$   \\
$c_{c}^{ZZ}$ & $-8.3\times 10^{-3}$ & $2.5\times 10^{-2}$   \\\hline
$d_{c}^{TT}$ & $-5.4\times 10^{-1}$ & $1.8\times 10^{-1}$   \\
$d_{c}^{ZZ}$ & $-9.2\times 10^{-3}$ & $2.8\times 10^{-2}$   \\\hline
$c_{s}^{TT}$ & \textcolor{gray}{$-1.3\times 10^1$} & \textcolor{gray}{$4.3$}   \\
$c_{s}^{ZZ}$ & $-1.7\times 10^{-1}$ & $5.2\times 10^{-1}$   \\\hline\hline
$c_{c}^{TT}$ & $-8.9\times 10^{-1}$ & $2.9\times 10^{-1}$   \\
$c_{c}^{ZZ}$ & $-5.6\times 10^{-2}$ & $1.7\times 10^{-1}$   \\\hline
$d_{b}^{TT}$ & $-9.5\times 10^{-1}$ & $3.1\times 10^{-1}$   \\
$d_{b}^{ZZ}$ & $-6.0\times 10^{-2}$ & $1.8\times 10^{-1}$   \\\hline\hline
\end{tabular}
\caption{Constraints on the $d=4$ coefficients that do not induce sidereal-time dependence. Constraints in gray do not satisfy the condition of perturbativity (coefficient $\lesssim 1$).\label{tab:MZindependent_cd}}
\end{table}

\def\arraystretch{1.2}
\begin{table}[t]
\begin{center}
\begin{tabular}{|c||c|c|c|}\hline\hline
coefficient [$\text{GeV}^{-1}$]   & $\delta_{\rm th}$ & $\delta_{\rm th}$, $\delta_{\rm lumi}$ &
  $\delta_{\rm th}$, $\delta_{\rm lumi}$, $\delta_{\rm syst}$ \\ \hline
$|a_{\text{S}u}^{(5)TXY}|$ & $4.7\times 10^{-8}$ & $1.7\times 10^{-8}$ & $5.3\times 10^{-10}$\\
$|a_{\text{S}u}^{(5)TXZ}|$ & $1.7\times 10^{-7}$ & $6.2\times 10^{-8}$ & $2.0\times 10^{-9}$\\
$|a_{\text{S}u}^{(5)TYZ}|$ & $1.7\times 10^{-7}$ & $6.3\times 10^{-8}$ & $2.0\times 10^{-9}$\\
$|a_{\text{S}u}^{(5)TXX}-a_{\text{S}u}^{(5)TYY}|$ & $3.0\times 10^{-7}$ & $1.1\times 10^{-7}$ & $3.4\times 10^{-9}$\\\hline
$|b_{\text{S}u}^{(5)TXY}|$ & $6.2\times 10^{-8}$ & $2.2\times 10^{-8}$ & $6.9\times 10^{-10}$\\
$|b_{\text{S}u}^{(5)TXZ}|$ & $2.2\times 10^{-7}$ & $8.2\times 10^{-8}$ & $2.6\times 10^{-9}$\\
$|b_{\text{S}u}^{(5)TYZ}|$ & $2.3\times 10^{-7}$ & $8.3\times 10^{-8}$ & $2.6\times 10^{-9}$\\
$|b_{\text{S}u}^{(5)TXX}-b_{\text{S}u}^{(5)TYY}|$ & $3.9\times 10^{-7}$ & $1.4\times 10^{-7}$ & $4.4\times 10^{-9}$\\\hline
$|a_{\text{S}d}^{(5)TXY}|$ & $2.5\times 10^{-6}$ & $8.8\times 10^{-7}$ & $2.8\times 10^{-8}$\\
$|a_{\text{S}d}^{(5)TXZ}|$ & $9.0\times 10^{-6}$ & $3.3\times 10^{-6}$ & $1.0\times 10^{-7}$\\
$|a_{\text{S}d}^{(5)TYZ}|$ & $9.2\times 10^{-6}$ & $3.3\times 10^{-6}$ & $1.0\times 10^{-7}$\\
$|a_{\text{S}d}^{(5)TXX}-a_{\text{S}d}^{(5)TYY}|$ & $1.6\times 10^{-5}$ & $5.6\times 10^{-6}$ & $1.8\times 10^{-7}$\\\hline\hline
$|a_{\text{S}c}^{(5)TXY}|$ & \textcolor{gray}{$1.0\times 10^{-3}$} & $3.7\times 10^{-4}$ & $1.2\times 10^{-5}$\\
$|a_{\text{S}c}^{(5)TXZ}|$ & \textcolor{gray}{$3.8\times 10^{-3}$} & \textcolor{gray}{$1.4\times 10^{-3}$} & $4.4\times 10^{-5}$\\
$|a_{\text{S}c}^{(5)TYZ}|$ & \textcolor{gray}{$3.9\times 10^{-3}$} & \textcolor{gray}{$1.4\times 10^{-3}$} & $4.4\times 10^{-5}$\\
$|a_{\text{S}c}^{(5)TXX}-a_{\text{S}c}^{(5)TYY}|$ & \textcolor{gray}{$6.6\times 10^{-3}$} & \textcolor{gray}{$2.4\times 10^{-3}$} & $7.5\times 10^{-5}$\\\hline
$|b_{\text{S}c}^{(5)TXY}|$ & \textcolor{gray}{$1.1\times 10^{-3}$} & $4.1\times 10^{-4}$ & $1.3\times 10^{-5}$\\
$|b_{\text{S}c}^{(5)TXZ}|$ & \textcolor{gray}{$4.2\times 10^{-3}$} & \textcolor{gray}{$1.5\times 10^{-3}$} & $4.8\times 10^{-5}$\\
$|b_{\text{S}c}^{(5)TYZ}|$ & \textcolor{gray}{$4.3\times 10^{-3}$} & \textcolor{gray}{$1.5\times 10^{-3}$} & $4.8\times 10^{-5}$\\
$|b_{\text{S}c}^{(5)TXX}-b_{\text{S}c}^{(5)TYY}|$ & \textcolor{gray}{$7.3\times 10^{-3}$} & \textcolor{gray}{$2.6\times 10^{-3}$} & $8.3\times 10^{-5}$\\\hline
$|a_{\text{S}s}^{(5)TXY}|$ & \textcolor{gray}{$2.1\times 10^{-2}$} & \textcolor{gray}{$7.6\times 10^{-3}$} & $2.4\times 10^{-4}$\\
$|a_{\text{S}s}^{(5)TXZ}|$ & \textcolor{gray}{$7.7\times 10^{-2}$} & \textcolor{gray}{$2.8\times 10^{-2}$} & $8.9\times 10^{-4}$\\
$|a_{\text{S}s}^{(5)TYZ}|$ & \textcolor{gray}{$7.9\times 10^{-2}$} & \textcolor{gray}{$2.8\times 10^{-2}$} & $8.9\times 10^{-4}$\\
$|a_{\text{S}s}^{(5)TXX}-a_{\text{S}s}^{(5)TYY}|$ & \textcolor{gray}{$1.4\times 10^{-1}$} & \textcolor{gray}{$4.8\times 10^{-2}$} & \textcolor{gray}{$1.5\times 10^{-3}$}\\\hline\hline
$|a_{\text{S}b}^{(5)TXY}|$ & \textcolor{gray}{$6.9\times 10^{-3}$} & \textcolor{gray}{$2.5\times 10^{-3}$} & $7.8\times 10^{-5}$\\
$|a_{\text{S}b}^{(5)TXZ}|$ & \textcolor{gray}{$2.5\times 10^{-2}$} & \textcolor{gray}{$9.2\times 10^{-3}$} & $2.9\times 10^{-4}$\\
$|a_{\text{S}b}^{(5)TYZ}|$ & \textcolor{gray}{$2.6\times 10^{-2}$} & \textcolor{gray}{$9.3\times 10^{-3}$} & $2.9\times 10^{-4}$\\
$|a_{\text{S}b}^{(5)TXX}-a_{\text{S}b}^{(5)TYY}|$ & \textcolor{gray}{$4.4\times 10^{-2}$} & \textcolor{gray}{$1.6\times 10^{-2}$} & $5.0\times 10^{-4}$\\\hline
$|b_{\text{S}b}^{(5)TXY}|$ & \textcolor{gray}{$7.4\times 10^{-3}$} & \textcolor{gray}{$2.6\times 10^{-3}$} & $8.4\times 10^{-5}$\\
$|b_{\text{S}b}^{(5)TXZ}|$ & \textcolor{gray}{$2.7\times 10^{-2}$} & \textcolor{gray}{$9.8\times 10^{-3}$} & $3.1\times 10^{-4}$\\
$|b_{\text{S}b}^{(5)TYZ}|$ & \textcolor{gray}{$2.8\times 10^{-2}$} & \textcolor{gray}{$1.0\times 10^{-2}$} & $3.1\times 10^{-4}$\\
$|b_{\text{S}b}^{(5)TXX}-b_{\text{S}b}^{(5)TYY}|$ & \textcolor{gray}{$4.7\times 10^{-2}$} & \textcolor{gray}{$1.7\times 10^{-2}$} & $5.3\times 10^{-4}$\\\hline\hline
\end{tabular}
\caption{Expected constraints on the $d=5$ coefficients that induce sidereal-time dependence, in units of $\text{GeV}^{-1}$. The part of the experimental uncertainties assumed to be 100\% correlated between binned data is indicated in the column labels. Constraints in gray do not satisfy the condition of perturbativity (coefficient $\lesssim E_p^{-1} \simeq 10^{-4}$\;GeV$^{-1}$).
\label{tab:MZ_ATLAS}}
\end{center}
\end{table}

\begin{table}[t]
\centering
\begin{tabular}{|c||c|c|c|}\hline\hline
coefficient  & $\delta_{\rm th}$ & $\delta_{\rm th}$, $\delta_{\rm lumi}$ &
  $\delta_{\rm th}$, $\delta_{\rm lumi}$, $\delta_{\rm syst}$ \\ \hline
$|c_{u}^{XY}|$ & $2.8\times 10^{-4}$ & $1.0\times 10^{-4}$ & $3.1\times 10^{-6}$ \\ 
$|c_{u}^{XZ}|$ & $9.4\times 10^{-4}$ & $3.4\times 10^{-4}$ & $1.1\times 10^{-5}$ \\ 
$|c_{u}^{YZ}|$ & $9.3\times 10^{-4}$ & $3.3\times 10^{-4}$ & $1.0\times 10^{-5}$ \\ 
$|c_{u}^{XX}-c_{u}^{YY}|$ & $2.0\times 10^{-3}$ & $7.2\times 10^{-4}$ & $2.3\times 10^{-5}$ \\ \hline
$|c_{d}^{XY}|$ & $1.2\times 10^{-2}$ & $4.4\times 10^{-3}$ & $1.4\times 10^{-4}$ \\ 
$|c_{d}^{XZ}|$ & $4.2\times 10^{-2}$ & $1.5\times 10^{-2}$ & $4.7\times 10^{-4}$ \\ 
$|c_{d}^{YZ}|$ & $4.1\times 10^{-2}$ & $1.5\times 10^{-2}$ & $4.6\times 10^{-4}$ \\ 
$|c_{d}^{XX}-c_{d}^{YY}|$ & $8.9\times 10^{-2}$ & $3.2\times 10^{-2}$ & $1.0\times 10^{-3}$ \\ \hline
$|d_{u}^{XY}|$ & $3.6\times 10^{-4}$ & $1.3\times 10^{-4}$ & $4.0\times 10^{-6}$ \\ 
$|d_{u}^{XZ}|$ & $1.2\times 10^{-3}$ & $4.3\times 10^{-4}$ & $1.4\times 10^{-5}$ \\ 
$|d_{u}^{YZ}|$ & $1.2\times 10^{-3}$ & $4.3\times 10^{-4}$ & $1.3\times 10^{-5}$ \\ 
$|d_{u}^{XX}-d_{u}^{YY}|$ & $2.6\times 10^{-3}$ & $9.3\times 10^{-4}$ & $2.9\times 10^{-5}$ \\ \hline\hline
$|c_{c}^{XY}|$ & $5.3\times 10^{-3}$ & $1.9\times 10^{-3}$ & $6.0\times 10^{-5}$ \\ 
$|c_{c}^{XZ}|$ & $1.8\times 10^{-2}$ & $6.5\times 10^{-3}$ & $2.0\times 10^{-4}$ \\ 
$|c_{c}^{YZ}|$ & $1.8\times 10^{-2}$ & $6.4\times 10^{-3}$ & $2.0\times 10^{-4}$ \\ 
$|c_{c}^{XX}-c_{c}^{YY}|$ & $3.8\times 10^{-2}$ & $1.4\times 10^{-2}$ & $4.4\times 10^{-4}$ \\ \hline
$|c_{s}^{XY}|$ & $1.1\times 10^{-1}$ & $3.9\times 10^{-2}$ & $1.2\times 10^{-3}$ \\ 
$|c_{s}^{XZ}|$ & $3.7\times 10^{-1}$ & $1.3\times 10^{-1}$ & $4.2\times 10^{-3}$ \\ 
$|c_{s}^{YZ}|$ & $3.6\times 10^{-1}$ & $1.3\times 10^{-1}$ & $4.0\times 10^{-3}$ \\ 
$|c_{s}^{XX}-c_{s}^{YY}|$ & $7.9\times 10^{-1}$ & $2.8\times 10^{-1}$ & $8.9\times 10^{-3}$ \\ \hline
$|d_{c}^{XY}|$ & $5.9\times 10^{-3}$ & $2.1\times 10^{-3}$ & $6.7\times 10^{-5}$ \\ 
$|d_{c}^{XZ}|$ & $2.0\times 10^{-2}$ & $7.2\times 10^{-3}$ & $2.3\times 10^{-4}$ \\ 
$|d_{c}^{YZ}|$ & $2.0\times 10^{-2}$ & $7.0\times 10^{-3}$ & $2.2\times 10^{-4}$ \\ 
$|d_{c}^{XX}-d_{c}^{YY}|$ & $4.3\times 10^{-2}$ & $1.5\times 10^{-2}$ & $4.8\times 10^{-4}$ \\ \hline\hline
$|c_{b}^{XY}|$ & $3.6\times 10^{-2}$ & $1.3\times 10^{-2}$ & $4.1\times 10^{-4}$ \\ 
$|c_{b}^{XZ}|$ & $1.2\times 10^{-1}$ & $4.4\times 10^{-2}$ & $1.4\times 10^{-3}$ \\ 
$|c_{b}^{YZ}|$ & $1.2\times 10^{-1}$ & $4.3\times 10^{-2}$ & $1.3\times 10^{-3}$ \\ 
$|c_{b}^{XX}-c_{b}^{YY}|$ & $2.6\times 10^{-1}$ & $9.3\times 10^{-2}$ & $2.9\times 10^{-3}$ \\ \hline
$|d_{b}^{XY}|$ & $3.8\times 10^{-2}$ & $1.4\times 10^{-2}$ & $4.3\times 10^{-4}$ \\ 
$|d_{b}^{XZ}|$ & $1.3\times 10^{-1}$ & $4.7\times 10^{-2}$ & $1.5\times 10^{-3}$ \\ 
$|d_{b}^{YZ}|$ & $1.3\times 10^{-1}$ & $4.6\times 10^{-2}$ & $1.4\times 10^{-3}$ \\ 
$|d_{b}^{XX}-d_{b}^{YY}|$ & $2.8\times 10^{-1}$ & $9.9\times 10^{-2}$ & $3.1\times 10^{-3}$ \\ \hline\hline
\end{tabular}
\caption{Expected constraints on the $d=4$ coefficients that induce sidereal-time dependence. The part of the experimental uncertainties assumed to be 100\% correlated between binned data is indicated in the column labels. 
\label{tab:MZ_ATLAS_cd}}
\end{table}

\section{Numerical results}
\label{sec:numerical}
The Drell-Yan cross section on the $Z$-boson pole has been measured by the ATLAS and CMS Collaborations at several center-of-mass energies: ATLAS (8 TeV, 20.2 $\text{fb}^{-1}$)~\cite{ATLAS:2023lsr}, ATLAS (13 TeV, 36.1 $\text{fb}^{-1}$)~\cite{ATLAS:2019zci}, CMS (13 TeV, 35.9 $\text{fb}^{-1}$)~\cite{Sirunyan:2019bzr} and ATLAS (13.6 TeV, 29 $\text{fb}^{-1}$)~\cite{ATLAS:2024irg}. All measurements have a total uncertainty of around 2\% and are compatible with each other and with the corresponding SM predictions at Next-to-Next-to-Leading-Order (NNLO)~\cite{Anastasiou:2003ds, Catani:2007vq, Catani:2009sm, Gavin:2010az, Gavin:2012sy, Bardin:2012jk, Li:2012wna, Arbuzov:2012dx}. Using these measurements, we assess the sensitivity to the $d=5$ coefficients and update the analysis of the renormalizable and CPT-even $c$- and $d$-type coefficients first studied in Ref.~\cite{Lunghi:2020hxn}. 

We extract actual constraints on those mass eigenstate coefficients whose effects do not vanish upon averaging over time. Imposing the condition of a vanishing trace, these coefficients are $a_{\text{S}f}^{(5)TAB}$, $b_{\text{S}f}^{(5)TAB}$, $c_f^{AB}$ and $d_f^{AB}$, where $AB=TT,ZZ$. Expected constraints on the remaining coefficients ($AB = XY,\; XZ,\; YZ,\; XX-YY$) can be obtained by simulating a sidereal-time analysis. For definiteness, we focus on the 13 TeV ATLAS measurement~\cite{ATLAS:2019zci} with the understanding that the other three measurements will yield similar constraints. The measured~\cite{ATLAS:2019zci} and NNLO (as presented in Ref.~\cite{ATLAS:2016oxs}) cross sections read: 
\begin{align}
\sigma_{\rm exp} & \; = 
736.2 \pm 0.2_\text{stat} \pm 6.4_{\rm syst} \pm 15.5_{\rm lumi} \; \text{pb} \; ,  \\
\sigma_{\rm NNLO} & \; = 
703^{+19}_{-24}{}^{+6}_{-8} {}^{+4}_{-6} {}^{+5}_{-5} \; \text{pb} \; .
\end{align} 
Constraints on the SME coefficients can be extracted from the Lorentz-violating part of the cross section evaluated on the $Z$-boson pole in the narrow-width approximation.  For $E_p = 6.5$ TeV this reads, with $\chi \approx 43.7\degree$ and $\psi \approx 168.7\degree$, 
\begin{align}
\label{totalxsec}
\sigma  & = 
6.3\times 10^8 \; a_{\text{S}Q_1}^{(5)000} +1.9\times 10^9 \; a_{\text{S}Q_1}^{(5)033} \nonumber\\
&+8.4\times 10^7 \; a_{\text{S}U_1}^{(5)000} +2.5\times 10^8 \; a_{\text{S}U_1}^{(5)033} \nonumber\\
&+6.7\times 10^6 \; a_{\text{S}D_1}^{(5)000} +2.0\times 10^7 \; a_{\text{S}D_1}^{(5)033}  \nonumber\\
&+9.3\times 10^4 \; a_{\text{S}Q_2}^{(5)000} -9.3\times 10^4 \; a_{\text{S}Q_2}^{(5)033} \nonumber\\
&+4.5\times 10^3 \; a_{\text{S}U_2}^{(5)000} -4.5\times 10^3 \; a_{\text{S}U_2}^{(5)033} \nonumber\\
&+2.3\times 10^3 \; a_{\text{S}D_2}^{(5)000} -2.3\times 10^3 \; a_{\text{S}D_2}^{(5)033}  \nonumber\\
 &+1.4\times 10^4 \; a_{\text{S}Q_3}^{(5)000} -1.4\times 10^4 \; a_{\text{S}Q_3}^{(5)033}\nonumber\\
 &+4.6\times 10^2 \; a_{\text{S}D_3}^{(5)000}  -4.6\times 10^2 \; a_{\text{S}D_3}^{(5)033}  \; ,
\end{align}
where the numerical coefficients are in units of pb$\cdot$GeV. As mentioned at the end of Sec.~\ref{sec:xsec}, sea-quark contributions are suppressed relative to the valence quark ones. For generality, the result~\eqref{totalxsec} is presented in the weak-eigenstate basis. It can be converted to the mass-eigenstate basis using Eqs.~\eqref{eq:quarkcoeffs1}-\eqref{eq:quarkcoeffs4}.

The extraction of constraints on coefficients with $TT$ and $ZZ$ indices, which do not induce sidereal oscillations, is straightforward: we simply require that the difference between the measured and SM cross sections (adding all uncertainties in quadrature) is saturated by the Lorentz-violating contribution. The resulting constraints at 95\% confidence level are presented in Tables~\ref{tab:MZindependent} and \ref{tab:MZindependent_cd}. Note that the small asymmetry between lower and upper constraints is due to the difference between the experimental and SM central values. There are no prior constraints reported on this subset of $d=5$ coefficients, so Table~\ref{tab:MZindependent} represents first constraints. We consider the results for the $d=4$ coefficients (Table~\ref{tab:MZindependent_cd}) as superseding Ref.~\cite{Lunghi:2020hxn}. For the $ZZ$-indexed coefficients, our results on $c_f^{ZZ}$ exceed those extracted using LEP data~\cite{Karpikov:2016qvq} by between one to two orders in magnitude. No previous constraints have been placed on $d_b^{TT}$ and $d_b^{ZZ}$, so these are also first constraints involving the bottom quark. 

Expected constraints on coefficients that induce sidereal oscillations of the cross section are calculated using the same strategy discussed in Ref.~\cite{Lunghi:2020hxn}. The basic idea is to simulate the result of a sidereal-time analysis in which all events are divided into $n$ sidereal bins. In Ref.~\cite{Lunghi:2020hxn}, we used the minimal choice $n=4$, but in an actual experimental search a much larger number of bins might be considered (for instance, in the DIS sidereal analysis of Ref.~\cite{ZEUS:2022msi} where $n=100$ was used). The crucial part of this simulated analysis is estimating the sidereal bin-to-bin correlation between systematic uncertainties (the statistical uncertainty simply increases by $\sqrt{n}$ and is uncorrelated). The sources of systematic uncertainties are theory ($\delta_{\rm th}$), luminosity ($\delta_{\rm lumi}$), and overall systematic ($\delta_{\rm syst}$). In regards to the analysis in Ref.~\cite{ATLAS:2019zci}, we considered three scenarios in which 
(1) $\delta_{\rm th}$, 
(2) $\delta_{\rm th}$ and $\delta_{\rm lumi}$,
and 
(3) $\delta_{\rm th}$, $\delta_{\rm lumi}$ and $\delta_{\rm syst}$ 
are assumed to be correlated in time. The results are presented in Tables~\ref{tab:MZ_ATLAS} and \ref{tab:MZ_ATLAS_cd}.

It is difficult to assess what fraction of the total uncertainty is correlated in time without a dedicated experimental analysis. Nonetheless, in Ref.~\cite{ATLAS:2019zci} explicit results where presented for the normalized distributions of the dilepton transverse momentum and of the angle $\phi^*_\eta$ (see Ref.~\cite{ATLAS:2019zci} for the exact definition). In both cases, the bin uncertainties are dominated by statistical fluctuations indicating that, for these two distributions, systematic and luminosity uncertainties are close to 100\% correlated across all bins (see Figure~3 of Ref.~\cite{ATLAS:2019zci}). If this feature persists for normalized distributions binned in sidereal time, only statistical uncertainties would be uncorrelated across sidereal bins and the expected constraints in the last columns of Tables~\ref{tab:MZ_ATLAS} and \ref{tab:MZ_ATLAS_cd} apply. 

Prior constraints on the $d=5$ coefficients inducing sidereal oscillations were placed by the ZEUS Collaboration~\cite{ZEUS:2022msi}. This search only extracted constraints on the $a_{\text{S}f}^{(5)}$-type coefficients with $f = u,d$. This is because in the DIS process with photon exchange, the CPT-odd property of the coefficients produces a cancellation when summing contributions from the sea quarks and antiquarks. All spin-dependent $b_{\text{S}f}^{(5)}$-type coefficients are also unobservable for this process due to the parity invariance of QED. Comparison with our expected results in Table~\ref{tab:MZ_ATLAS} thus suggests an ATLAS or CMS sidereal analysis could improve constraints on valence-quark coefficients by up to three orders in magnitude and probe effects involving sea quarks. It is worth noting that our expected valence-quark constraints are roughly the same level as those placed on effective $d=5$ proton coefficients from the 1S-2S transition in atomic hydrogen~\cite{Kostelecky:2015nma}. The expected constraints on the $d=4$ coefficients in Table~\ref{tab:MZ_ATLAS_cd} are within about an order of magnitude of the ZEUS results on the $c_{f}$-type coefficients with $f = u,d,s$~\cite{ZEUS:2022msi}. As the $d_{f}$-type coefficients are also spin-dependent, they are similarly unobservable in photon-mediated DIS. An ATLAS or CMS sidereal analysis could thus probe, for the first time, a number of sea-quark and spin-dependent Lorentz- and CPT-violating effects.

\bibliography{references}

\section{Acknowledgments}
NS acknowledges funding by the Deutsche Forschungsgemeinschaft (DFG) under the Heinz Maier Leibnitz Prize - BeyondSM\_HML-537662082.

\end{document}